\documentclass[prb,aps,eqsecnum]{revtex4}

\usepackage[dvips]{graphics}
\usepackage{psfig}
\usepackage{graphicx}
\usepackage{amssymb}
\usepackage{epstopdf}

\begin{document}


\title{Coherent effects in
double-barrier ferromagnet/superconductor/ferromagnet junctions}
\author{Milo\v{s} Bo\v{z}ovi\'{c} and Zoran Radovi\'c}
\address{Department of Physics, University of Belgrade, P.O. Box 368, 11001 Belgrade, Serbia and Montenegro}

\begin{abstract}
Coherent quantum transport in
ferromagnet/superconductor/ferromagnet (FSF) double-barrier
junctions is studied. Analytic expressions for charge and spin
conductance spectra are derived for the general case of insulating
interfaces (from metallic to tunnel limit), the Fermi velocity
mismatch, and for parallel (P) and antiparallel (AP) alignment of
the electrode magnetizations. We focus on two characteristic
features of finite size and coherency: subgap electronic
transport, and oscillations of the differential conductance.
Periodic vanishing of the Andreev reflection at the energies of
geometrical resonances above the superconducting gap is a striking
consequence of the quasiparticle interference. In contrast with
the case of incoherent transport, a non-trivial spin-polarization
without the excess spin accumulation is found for the AP
alignment.
\end{abstract}


\maketitle
\section{Introduction}

During the past decade, there has been a growing interest in
various electronic systems driven out of equilibrium by the
injection of spin-polarized carriers. Such systems can be realized
by current-biasing structures consisting of ferromagnetic and
nonferromagnetic (e.g., superconducting) layers, due to the
difference in population of majority- and minority- spin
subbands.\cite{Prinz} The concept of spin-polarized current
nowadays has attracted considerable interest in ferromagnetic
heterostructures, in particular for applications in
spintronics.\cite{Osofsky}

Charge transport through a normal metal/superconductor (NS)
junction, with an insulating barrier of arbitrary strength at the
interface, has been studied by Blonder, Tinkham, and Klapwijk
(BTK),\cite{BTK} and the Andreev reflection is recognized as the
mechanism of normal-to-supercurrent
conversion.\cite{Andreev,Furusaki Tsukada} The BTK theory has been
extended by Tanaka and Kashiwaya to include the anisotropy of the
pair potential in $d$-wave superconductors.\cite{Tanaka 95,Tanaka
00} The modification of the Andreev reflection by the spin
injection from a ferromagnetic metal into a superconductor in
ferromagnet/superconductor (FS) junctions was first analyzed by de
Jong and Beenakker.\cite{dJB} More recently, the effects of
unconventional $d$-wave and $p$-wave pairing and of the exchange
interaction in FS systems, such as the zero-bias conductance peak
and the virtual Andreev reflection, have been clarified by
Kashiwaya {\it et al}.\cite{Beasley} and Yoshida {\it et
al}.\cite{Yoshida} The Fermi velocity mismatch between two metals
can also significantly affect the Andreev reflection by altering
the subgap conductance,\cite{Zutic} which is similar to the
presence of an insulating barrier.\cite{Zhu}

In experiments, a superconductor is used to determine the spin
polarization of the current injected from (or into) a ferromagnet
by measuring the differential conductance. These measurements have
been performed on tunnel junctions in an external magnetic
field,\cite{Tedrow,Platt} metallic point
contacts,\cite{Soulen,Novo} nano-contacts formed by
microlithography,\cite{Upad} and FS junctions with $d$-wave
superconductors, grown by molecular beam epitaxy.\cite{Vasko} In
diffusive FS junctions, the excess resistance may be induced by
spin accumulation near the insulating interface\cite{Jedema} and
by the proximity effect.\cite{Gueron,Petrashov,Sillanpaa}

When interfaces act incoherently, the BTK model can be
successfully applied to normal metal/superconductor/normal metal
(NSN) or ferromagnet/superconductor/ferromagnet (FSF) double
junctions.\cite{Takahashi,Zheng} However, the properties of
coherent quantum transport in clean superconducting
heterostructures are strongly influenced by size effects, which
are not included in the BTK model. Well-known examples are the
current-carrying Andreev bound states\cite{Tanaka 00,Nazarov} and
multiple Andreev reflections\cite{KBT,Basel,Ingerman,Brinkman} in
superconductor/normal metal/superconductor (SNS) junctions. Since
early experiments by Tomasch,\cite{Tomasch} the geometric
resonance nature of the differential conductance oscillations in
SNS and NSN tunnel junctions has been ascribed to the electron
interference in the central
film.\cite{Anderson,Rowell,McMillan,Kanadjani} Recently,
McMillan-Rowell oscillations were observed in SNS edge junctions
of $d$-wave superconductors and used for measurements of the
superconducting gap and the Fermi velocity.\cite{Nesher}

In this paper we study coherent electronic transport in FSF
double-barrier junctions (and NSN as a special case) within the
framework of BCS theory. We limit ourselves to clean conventional
(isotropic and $s$-wave) superconductors, and neglect, for
simplicity, the self-consistency of the pair
potential\cite{Geers,Buzdin} and nonequlibrium effects of charge
and spin accumulation at the interfaces.\cite{FalkoB,McCann} When
two interfaces are recognized by electrons simultaneously,
characteristic features of finite size and coherency are the
subgap transport of electrons and oscillations of both charge and
spin differential conductances above the gap. One consequence of
the quasiparticle interference is the periodic vanishing of the
Andreev reflection at the energies of geometrical resonances. The
other is the existence of a nontrivial spin polarization of the
current not only for the parallel (P), but also for the
antiparallel (AP) alignment of the electrode magnetizations.
Previous analysis of incoherent transport in FSF double junctions
in AP alignment predict the absence of spin current and
suppression of superconductivity with increasing voltage, as a
result of spin imbalance in the superconducting
film.\cite{Takahashi,Zheng}

\section{Scattering probabilities}

We consider an FSF double junction consisting of a clean
superconducting layer of thickness $l$, connected to ferromagnetic
electrodes by thin, insulating interfaces. For the ferromagnetic
metal we adopt the Stoner model, describing the spin-polarization
effect by the usual one-electron Hamiltonian with an exchange
potential. The quasiparticle propagation is described by the
Bogoliubov--de Gennes equation
\begin{eqnarray}
\left(
\begin{array}{ccc}
  H_0({\bf r})-\rho_{\sigma}h({\bf r}) && \Delta({\bf r}) \\
  \Delta^{*}({\bf r}) && -H_0({\bf r})+\rho_{\bar{\sigma}}h({\bf
r})
\end{array}
\right) \Psi_\sigma({\bf r})~=~E\Psi_\sigma({\bf r}), \label{BdG}
\end{eqnarray}
with $H_{0}({\bf r})=-\hbar^{2}\nabla^{2}/2m+W({\bf r})+U({\bf
r})-\mu$, where $U({\bf r})$ and $\mu$ are the Hartree and the
chemical potential, respectively. The interface potential is
modeled by $W({\bf r})=\hat{W}\{\delta(z)+\delta(z-l)\}$, where
the $z$ axis is perpendicular to the layers and $\delta(z)$ is the
Dirac $\delta$ function. Neglecting the self-consistency of the
superconducting pair potential, $\Delta({\bf r})$ is taken in the
form $\Delta \Theta(z) \Theta(l-z)$, where $\Theta(z)$ is the
Heaviside step function and $\Delta$ is the bulk superconducting
gap. In Eq. (\ref{BdG}), $\sigma$ is the quasiparticle spin
($\sigma =\uparrow ,\downarrow$ and $\bar{\sigma}=\downarrow
,\uparrow$), $E$ is the energy with respect to $\mu$, $h({\bf r})$
is the exchange potential given by
$h_{0}\{\Theta(-z)+[-]\Theta(z-l)\}$ for the P [AP] alignment, and
$\rho_{\sigma}$ is $1$ ($-1$) for  spins up (down). The electron
effective mass $m$ is assumed to be the same for the whole
junction. Here, $\mu-U({\bf r})$ is the Fermi energy of the
superconductor, $E^{(S)}_F$, or the mean Fermi energy of a
ferromagnet, $E^{(F)}_F=(E^\uparrow_F+E^\downarrow_F)/2$. Moduli
of the Fermi wave vectors, $k^{(F)}_F= \sqrt{2mE^{(F)}_F/\hbar^2}$
and $k^{(S)}_F= \sqrt{2mE^{(S)}_F/\hbar^2}$, can be different in
general, and in the following, the Fermi wave vector mismatch
(FWVM) will be taken into account through the parameter
$\kappa=k^{(F)}_F/k^{(S)}_F$. The parallel component of the wave
vector ${\bf k}_{||,\sigma}$ is conserved, and the wave function
\begin{equation}
\Psi_\sigma({\bf r})=\exp(i{\bf k}_{||,\sigma} \cdot {\bf
r})~\psi_\sigma(z)
\end{equation}
satisfies appropriate boundary conditions. Four independent
solutions of Eq. (\ref{BdG}) correspond to the four types of
injection: an electron or a hole from either the left or from the
right electrode.\cite{Furusaki Tsukada}

For the injection of an electron from the left, with energy $E>0$,
spin $\sigma$, and angle of incidence $\theta$ (measured from the
$z$ axis), solution for $\psi_\sigma (z)$ in various regions has
the following form:

\noindent in the left ferromagnet ($z<0$)
\begin{equation} \psi_\sigma
(z)=\{\exp(ik^+_{\sigma}
z)+b_{\sigma}(E,\theta)\exp(-ik^+_{\sigma}
z)\}\left(\begin{array}{c}
     1 \\
     0 \\
   \end{array}\right)+
    a_{\sigma}(E,\theta)\exp(ik^-_{\bar{\sigma}} z)\left(\begin{array}{c}
      0 \\
      1 \\
    \end{array}\right);
\label{psiL}
\end{equation}
in the superconductor ($0<z<l$),
\begin{eqnarray}
\psi_\sigma (z)&=&\{
c_{1}(E,\theta)\exp(iq^+_{\sigma}z)+c_{2}(E,\theta)\exp(-iq^+_{\sigma}z)\}
\left(\begin{array}{c}
    \bar{u} \\
    \bar{v}
  \end{array}\right) \nonumber \\
&~&+\{
c_{3}(E,\theta)\exp(iq^-_{\sigma}z)+c_{4}(E,\theta)\exp(-iq^-_{\sigma}z)\}
\left(\begin{array}{c}
    \bar{v}^{*} \\
    \bar{u}^{*}
  \end{array}\right);
\label{psiS}
\end{eqnarray}
and in the right ferromagnet ($z>l$), for the P [AP] alignment of
the magnetizations,
\begin{equation}
 \psi_\sigma(z)=c_{\sigma}(E,\theta)\exp(ik^+_{\sigma[{\bar\sigma}]}z)\left(\begin{array}{c}
     1 \\
     0 \\
   \end{array}\right)+
    d_{\sigma}(E,\theta)\exp(-ik^-_{{\bar\sigma}[\sigma]}z)\left(\begin{array}{c}
      0 \\
      1 \\
    \end{array}\right).
\label{psiR}
\end{equation}
Here, $\bar{u}=\sqrt{(1+\Omega/E)/2}$ and
$\bar{v}=\sqrt{(1-\Omega/E)/2}$ are the BCS coherence factors, and
$\Omega=\sqrt{E^2-\Delta^2}$. The $z$ components of the wave
vectors are
\[
    k^\pm_{\sigma}=\sqrt{(2m/\hbar ^2)(E^{(F)}_F+\rho_{\sigma}h_0 \pm
    E)-{\bf k}^2_{||,\sigma}}
\]
and
\[
    q^\pm_{\sigma}=\sqrt{(2m/\hbar ^2)(E^{(S)}_F\pm\Omega)-{\bf
    k}^2_{||,\sigma}},
\]
where $|{\bf k}_{||,\sigma}|=\sqrt{(2m/\hbar
^2)(E^{(F)}_F+\rho_{\sigma}h_0+E)}~\sin\theta$. The coefficients
$a_{\sigma}$, $b_{\sigma}$, $c_{\sigma}$, and $d_{\sigma}$ are,
respectively, the probability amplitudes of (1) Andreev reflection
as a hole of the opposite spin (AR), (2) normal reflection as an
electron (NR), (3) transmission to the right electrode as an
electron (TE), and (4) transmission to the right electrode as a
hole of the opposite spin (TH). Processes (1) and (4) are
equivalent to the formation of a Cooper pair in the superconductor
by taking one more electron from either the left or the right
electrode, respectively. Amplitudes of the Bogoliubov electronlike
and holelike quasiparticles, propagating in the superconducting
layer, are given by the coefficients $c_1$ through $c_4$.

Neglecting small terms $E/E^{(F)}_F\ll 1$ and $\Delta/E^{(S)}_F\ll
1$ in the wave vectors, except in the exponents
\begin{equation}
\label{zeta} \zeta_\pm =l\left(q^+_{\sigma}\pm
q^-_{\sigma}\right),
\end{equation}
solutions for the probability amplitudes can be written in the
compact form for the general case (see the Appendix). In the
following we use the approximated wave-vector components in units
of $k^{(S)}_F$: $\tilde{q}_\sigma=\sqrt{1-\tilde{\bf
k}^2_{||,\sigma} }$, $\tilde{k}_{\sigma}=\lambda_\sigma
\cos\theta$ and $|\tilde{\bf k}_{||,\sigma}|=\lambda_\sigma
\sin\theta$, where $\lambda_\sigma=\kappa\sqrt{1+\rho_\sigma X}$,
$X=h_0 /E^{(F)}_F\geq 0$, and $\kappa\neq 1$ is measuring FWVM.
Dimensionless parameter measuring the strength of each interface
barrier is $Z={2m\hat{W}}/\hbar^2 k^{(S)}_F$.

Solutions for the other three types of injection can be obtained
by the same procedure. In particular, if a hole with energy $-E$,
spin $\sigma$, and angle of incidence $\theta$ is injected from
the left, the substitution $q^+_{\sigma} \rightleftharpoons
q^-_{\sigma}$ holds, and the scattering probabilities are the same
as for the injection of an electron with $E$, $\sigma$, and
$\theta$. Therefore, in order to include the description of both
electron and hole injections, the calculated probabilities should
be regarded as even functions of $E$. Also, for an electron or a
hole, injected from the right, the probabilities are the same as
for the injection from the left, except $\sigma\to\bar{\sigma}$
for the AP alignment.

From the probability current conservation, the probabilities of
outgoing particles satisfy the normalization condition
\begin{equation}
\label{ABCD}
A_\sigma(E,\theta)+B_\sigma(E,\theta)+C_\sigma(E,\theta)+D_\sigma(E,\theta)=1,
\end{equation}
where,
\begin{eqnarray}
\label{maliA}
A_\sigma(E,\theta)&=&\Re\left(\frac{\tilde{k}_{\bar{\sigma}}}
{\tilde{k}_{\sigma}}\right)|a_\sigma (E,\theta)|^2, \\
\label{maliB} B_\sigma(E,\theta)&=&|b_\sigma (E,\theta)|^2, \\
\label{maliC}
C_\sigma(E,\theta)&=&\Re\left(\frac{\tilde{k}_{\sigma[{\bar{\sigma}}]}}{\tilde{k}_{\sigma}}\right)
|c_\sigma(E,\theta)|^2, \\ \label{maliD} D_\sigma(E,\theta)&=&
\Re\left(\frac{\tilde{k}_{{\bar{\sigma}}[\sigma]}}{\tilde{k}_{\sigma}}\right)
|d_\sigma(E,\theta)|^2.
\end{eqnarray}

It follows from the general solution that
$A_\sigma(E,\theta)=D_\sigma(E,\theta)=0$ when
\begin{equation}
\label{resonance}
\zeta_-=2n\pi
\end{equation}
for $n=0,\pm 1,\pm 2,\ldots$, independently of $X$, $Z$, and
$\kappa$. Therefore, both direct and crossed Andreev reflection
vanishes at the energies of geometrical resonances in
quasiparticle spectrum. The absence of AR and TH processes means
that all quasiparticles with energies satisfying Eq.
(\ref{resonance}) will pass unaffected from one electrode to
another, without creation or annihilation of Cooper pairs. The
effect is similar to the over-the-barrier resonances in the simple
problem of one-particle scattering against a step-function
potential,\cite{CohenT} the superconducting gap playing the role
of a finite-width barrier.\cite{Tinkham}

\begin{figure}[h]
\begin{center}
    \includegraphics[width=10cm]{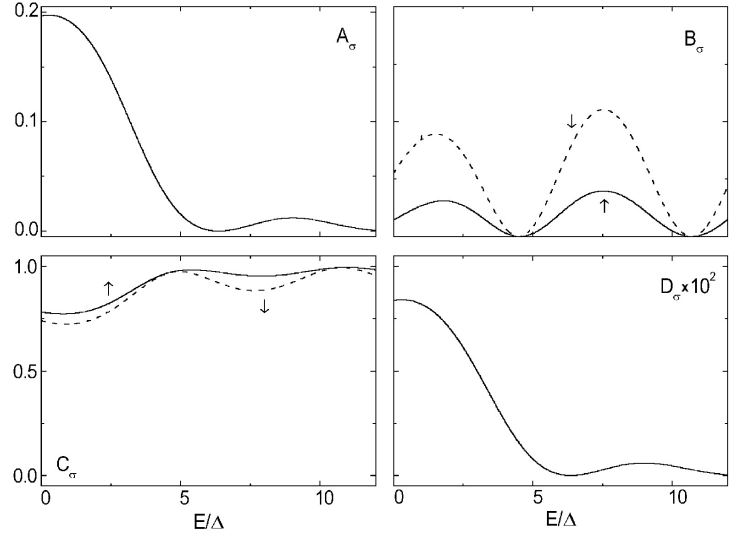}
    \caption{Scattering probabilities $A_\sigma(E,0)$,
            $B_\sigma(E,0)$, $C_\sigma(E,0)$, and $D_\sigma(E,0)$ for an FSF
            double junction with thin superconducting film, $lk^{(S)}_F=10^3$,
            for $X=0.5$, $Z=0$, $\kappa=1$, $\Delta/E^{(S)}_F=10^{-3}$, and P
            alignment. Solid curves: injection of an electron with
            $\sigma=\uparrow$. Dashed curves: injection of an electron with
            $\sigma=\downarrow$. Here, $A_\sigma(E,0)$ and $D_\sigma(E,0)$ are
            spin independent due to the singlet-state pairing and transparent
            interfaces (Ref.~48).}
    \label{thin}
\end{center}
\end{figure}

\begin{figure}[h]
\begin{center}
    \includegraphics[width=10cm]{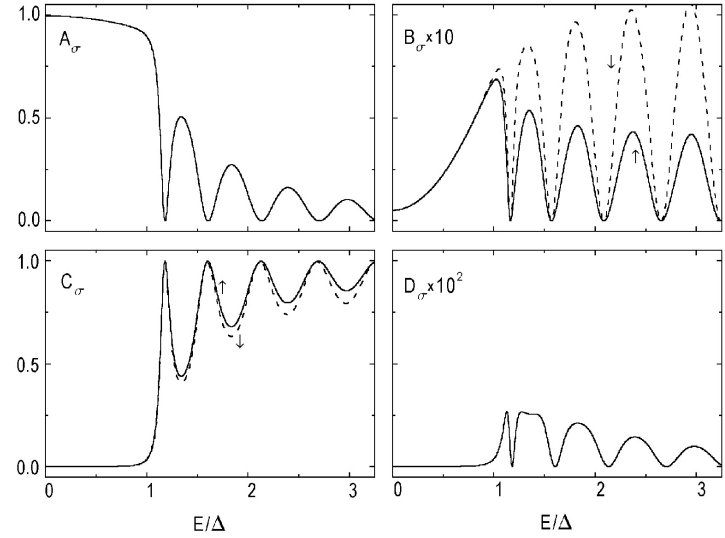}
    \caption{Scattering probabilities $A_\sigma(E,0)$,
            $B_\sigma(E,0)$, $C_\sigma(E,0)$, and $D_\sigma(E,0)$ for an FSF
            double junction with thick superconducting film,
            $lk^{(S)}_F=10^4$, for $X=0.5$, $Z=0$, $\kappa=1$,
            $\Delta/E^{(S)}_F=10^{-3}$, and P alignment. Solid curves:
            injection of an electron with $\sigma=\uparrow$. Dashed curves:
            injection of an electron with $\sigma=\downarrow$. Here,
            $A_\sigma(E,0)$ and $D_\sigma(E,0)$ are spin independent due to
            the singlet-state pairing and transparent interfaces
            (Ref.~48).}
    \label{thick}
\end{center}
\end{figure}

Characteristic features of coherent electronic transport through
the superconducting layer are the subgap transmission (without
conversion to supercurrent) and oscillations of the scattering
probabilities. For $E<\Delta$, the subgap transmission of
electrons or holes suppresses the Andreev reflection. For
$E>\Delta$, all probabilities oscillate with $E$ and $l$ due to
the interference of incoming and outgoing particles. These effects
are illustrated in Figs.~\ref{thin} and \ref{thick} for an FSF
double junction, with $Z=0$ and $\kappa=1$, in P alignment. Taking
$\Delta/E^{(S)}_F=10^{-3}$, in a thin superconducting film,
$lk^{(S)}_F\sim 10^3$, the Andreev reflection is strongly
suppressed, since the subgap transmission of electrons is
considerable, Fig.~{\ref{thin}}. In this case, the oscillations
are less pronounced, with the period much larger than $\Delta$.
For a thick film, $lk^{(S)}_F\sim 10^4$, the subgap tunneling is
irrelevant [except for small "tails" in $A_\sigma(E,0)$ and
$C_\sigma(E,0)$ at $E\lesssim\Delta$], and the oscillations above
the gap are more pronounced with the period on the order of
$\Delta$, Fig.~{\ref{thick}}. The scattering probabilities for AP
and P alignment differ very slightly in the case of normal
incidence, $\theta=0$. Although spin independent, $A_\sigma(E,0)$
is even more suppressed due to the exchange interaction. In
contrast with NSN junction with transparent interfaces,
$B_\sigma(E,0)$ and $D_\sigma(E,0)$ become nontrivial. The
spin-dependent $B_\sigma(E,0)$ has zeros at the same energies as
$A_\sigma(E,0)$ and $D_\sigma(E,0)$, so that maxima in
$C_\sigma(E,0)$ at geometrical resonances are still equal to unity
due to the interface transparency.

The insulating barriers ($Z>0$) and FWVM ($\kappa\neq 1$) reduce
AR and TE and enhance NR and TH probabilities, as well as the
exchange interaction ($X>0$). In contrast to the positions of
zeros of $A_\sigma(E,0)$, given by Eq. (\ref{resonance}), the
positions of maxima of $A_\sigma(E,0)$, as well as that of zeros
and maxima of $B_\sigma(E,0)$, $C_\sigma(E,0)$, and
$D_\sigma(E,0)$, are dependent on $X$, $Z$, and $\kappa$.
Approaching the tunnel limit ($Z\to\infty$), peaks in the
scattering probabilities gradually split into two spikes belonging
to consecutive pairs with positions defined by the quantization
conditions
\begin{equation}
\label{n} lq^+_{\sigma}=n_1\pi,~~~lq^-_{\sigma}=n_2\pi,
\end{equation}
giving the bound-state energies of an isolated superconducting
film. Eqs. (\ref{resonance}) and (\ref{n}) are simply connected by
$n_1-n_2=2n$. The exception is the spike at the gap edge,
originating from the singularity in the BCS density of states.
This gives the correspondence between the electronic transport
through resonances in metallic junctions and through bound states
in the tunnel junctions.

\section{Differential conductances}

When voltage $V$ is applied to the junction symmetrically, the
charge current density can be written in the
form\cite{Dong,Yamashita,Lambert}
\begin{eqnarray*}
j_q(V)&=&\frac{e{k^{(S)}_{F}}^2}{2\pi
h}\int\limits_{-\infty}^{\infty}{\rm d}E
\sum_{\sigma=\uparrow,\downarrow}\lambda^2_\sigma
\int\limits_{0}^{\pi/2}{\rm d}\theta \sin\theta \cos\theta
\left[1+A_\sigma(E,\theta)-B_\sigma(E,\theta)+C_\sigma(E,\theta)-D_\sigma(E,\theta)\right]\delta
f({\bf
k},V)\\
&=&\frac{e{k^{(S)}_{F}}^2}{\pi h}\int\limits_{-\infty}^{\infty}{\rm d}E
\sum_{\sigma=\uparrow,\downarrow}\lambda^2_\sigma
\int\limits_{0}^{\pi/2}{\rm d}\theta \sin\theta \cos\theta
\left[A_\sigma(E,\theta)+C_\sigma(E,\theta)\right]\delta f({\bf
k},V),
\end{eqnarray*}
where $\delta f({\bf k},V)$ is the asymmetric part of the
nonequilibrium distribution function of current carriers. In the last equality the normalization condition, Eq. (\ref{ABCD}), was taken into
account. Without
solving the suitable transport equation, we take $\delta f({\bf
k},V)=f_0(E-eV/2)-f_0(E+eV/2)$, where $f_0(E)$ is the Fermi-Dirac
equilibrium distribution function.\cite{BTK,Tinkham} In this
approach, the charge current per orbital transverse
channel is given by
\begin{equation}
\label{Iq} I_q(V)=\frac{1}{e}\int\limits_{-\infty}^{\infty}{\rm d}E
\left[f_0(E-eV/2)-f_0(E+eV/2)\right]G_q(E),
\end{equation}
where the differential charge conductance at zero temperature is
\begin{equation}
\label{3D q} G_q(E) =
\frac{e^2}{h}\sum_{\sigma=\uparrow,\downarrow}
    \lambda_\sigma^2 \int_0^{\pi/2} {\rm d}\theta\sin\theta\cos\theta
    ~\left[A_\sigma(E,\theta)+C_\sigma(E,\theta)\right].
\end{equation}

On the other hand, the probability current per orbital transverse
channel is given by
\begin{eqnarray*}
j_s(V)&=&\frac{e{k^{(S)}_{F}}^2}{2\pi
h}\int\limits_{-\infty}^{\infty}{\rm d}E
\sum_{\sigma=\uparrow,\downarrow}\rho_\sigma\lambda^2_\sigma
\int\limits_{0}^{\pi/2}{\rm d}\theta \sin\theta \cos\theta
\left[1-A_\sigma(E,\theta)-B_\sigma(E,\theta)+C_\sigma(E,\theta)+D_\sigma(E,\theta)\right]\delta
f({\bf
k},V)\\
&=&\frac{e{k^{(S)}_{F}}^2}{\pi h}\int\limits_{-\infty}^{\infty}{\rm d}E
\sum_{\sigma=\uparrow,\downarrow}\rho_\sigma\lambda^2_\sigma
\int\limits_{0}^{\pi/2}{\rm d}\theta \sin\theta \cos\theta
\left[1-A_\sigma(E,\theta)-B_\sigma(E,\theta)\right]\delta f({\bf
k},V).
\end{eqnarray*}
The corresponding spin current is then
\begin{equation}
\label{Is}
I_s(V)=\frac{1}{e}\int\limits_{-\infty}^{\infty}{\rm d}E\left[f_0(E-eV/2)-f_0(E+eV/2)\right]G_s
(E),
\end{equation}
where the differential spin conductance at zero temperature is
\begin{eqnarray}
\label{3D s}
G_s(E)&=&\frac{e^2}{h}\sum_{\sigma=\uparrow,\downarrow}\rho_\sigma
\lambda^2_\sigma \int\limits_{0}^{\pi/2}{\rm
d}\theta~\sin\theta\cos\theta
\left[1-A_\sigma(E,\theta)-B_\sigma(E,\theta)\right]\nonumber\\
&=&\frac{e^2}{h}\sum_{\sigma=\uparrow,\downarrow}\rho_\sigma
\lambda^2_\sigma \int\limits_{0}^{\pi/2}{\rm
d}\theta~\sin\theta\cos\theta
\left[C_\sigma(E,\theta)+D_\sigma(E,\theta)\right].
\end{eqnarray}

The upper limit of integration over $\theta$ is determined by
total reflection. Following the conservation of ${\bf
k}_{||,\sigma}$, transmission of an electron (hole) with
$\sigma=\uparrow$, injected from the left electrode into the
superconductor, is possible only for angles of incidence $\theta$
satisfying $\theta <\theta _{c1}$, where $\theta
_{c1}=\arcsin(1/\lambda_\uparrow)$ is the angle of total
reflection. Then, $A_\uparrow(E,\theta)=0$ and
$B_\uparrow(E,\theta)=1$ for $\theta >\theta _{c1}$. On the other
hand, $\tilde{k}_{\downarrow}$, which corresponds to the hole
(electron) created by the Andreev reflection, is real only for
$\theta<\theta_{c2}=
\arcsin(\lambda_\downarrow/\lambda_\uparrow)$. The virtual Andreev
reflection occurs for $\theta _{c2}<\theta<\theta _{c1}$, since
$\tilde{k}_{\downarrow}$ becomes imaginary in that
case.\cite{Beasley} For injection of an electron (hole) with
$\sigma=\downarrow$, transmission into the superconductor is
possible for any $\theta<\pi/2$, and $\tilde{k}_{\uparrow}$ is
always real.

\begin{figure}[h]
\begin{center}
    \includegraphics[width=10cm]{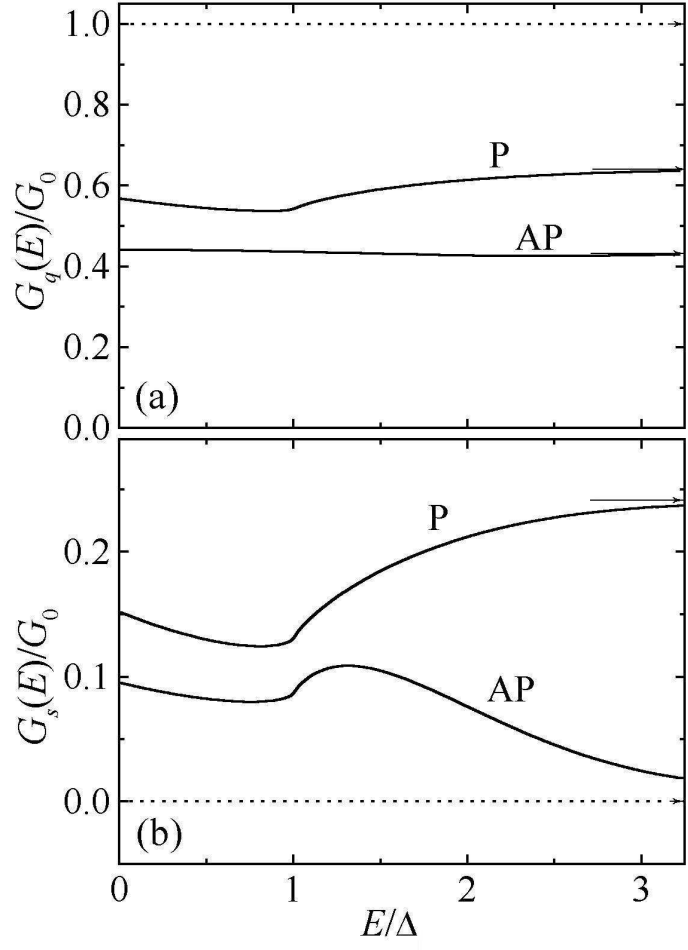}
    \caption{Differential charge (a) and spin (b)
            conductance spectra $G_q(E)$ and $G_s(E)$ of an FSF double planar
            junction with thin superconducting film, $lk^{(S)}_F=10^3$, for
            $X=0.5$, $Z=0$, $\kappa=1$, $\Delta/E^{(S)}_F=10^{-3}$, in P and
            AP alignment. Conductances of the corresponding NSN junction are
            shown for comparison (dotted lines). Arrows indicate $G^N_q$ and
            $G^N_s$ values. Here, $G_0=2e^2/h$ is the conductance quantum.}
    \label{l3}
\end{center}
\end{figure}

\begin{figure}[h]
\begin{center}
    \includegraphics[width=10cm]{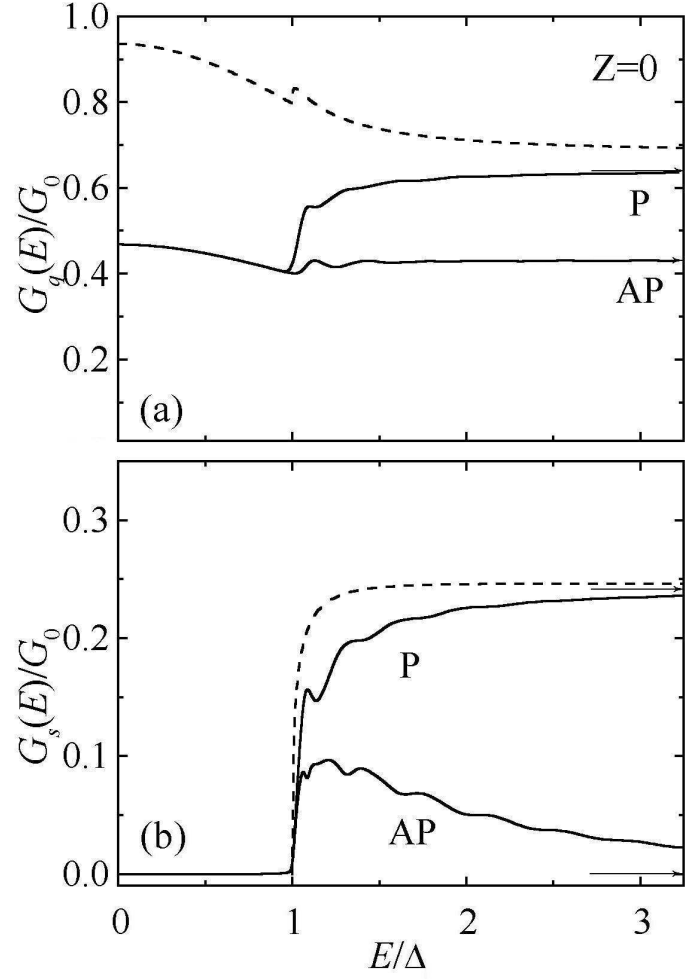}
    \caption{Differential charge (a) and spin (b)
            conductance spectra $G_q(E)$ and $G_s(E)$ of an FSF double planar
            junction with thick superconducting film, $lk^{(S)}_F=10^4$, for
            $X=0.5$, $Z=0$, $\kappa=1$, $\Delta/E^{(S)}_F=10^{-3}$, in P and
            AP alignment. Dashed curves represent the generalized BTK results
            for the same parameters. Arrows indicate $G^N_q$ and $G^N_s$
            values. Here, $G_0=2e^2/h$ is the conductance quantum.}
    \label{GZ0}
\end{center}
\end{figure}

\begin{figure}[h]
\begin{center}
    \includegraphics[width=10cm]{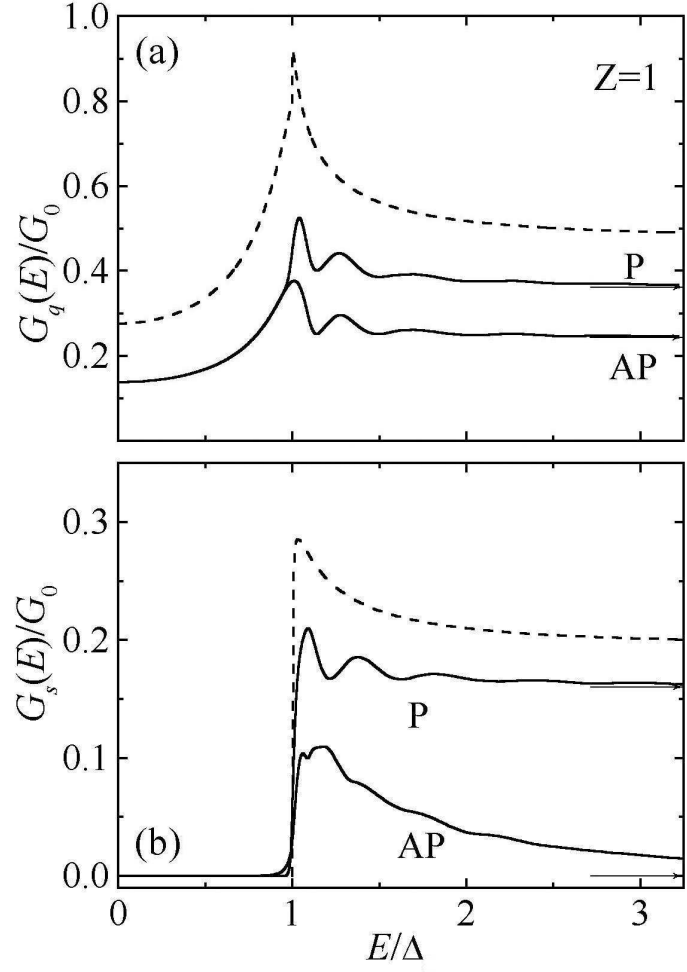}
    \caption{Differential charge (a) and spin (b)
            conductance spectra $G_q(E)$ and $G_s(E)$ of an FSF double planar
            junction with thick superconducting film, $lk^{(S)}_F=10^4$, for
            $X=0.5$, $Z=1$, $\kappa=1$, $\Delta/E^{(S)}_F=10^{-3}$, in P and
            AP alignment. Dashed curves represent the generalized BTK results
            for the same parameters. Arrows indicate $G^N_q$ and $G^N_s$
            values. Here, $G_0=2e^2/h$ is the conductance quantum.}
    \label{GZ1}
\end{center}
\end{figure}

The influence of the exchange interaction on the conductance
spectra is illustrated for $X=0.5$ and $Z=0$, for thin
(Fig.~\ref{l3}) and thick (Fig.~\ref{GZ0}) superconducting films.
The spin-polarized subgap transmission of quasiparticles, and
strong suppression of the Andreev reflection as a consequence is
significant in thin superconducting films,\cite{ft2} whereas the
conductance oscillations above the gap are pronounced in the thick
films. The magnetoresistance is apparent, as charge and spin
conductances are larger for the P than for the AP alignment. The
effect of interface resistance for weak nontransparency ($Z=1$) is
illustrated in Fig.~\ref{GZ1}. It can be shown that the FWVM has
an influence similar to the nontransparency, due to the
enhancement of normal reflection. Suppression of the conductance
is more significant for $\kappa>1$.

In Figs.~\ref{l3}--\ref{GZ1} we have indicated the values of
normal conductances $G^N_q$ and $G^N_s$ of the corresponding FNF
double planar junction, obtained by setting $A_\sigma(E,\theta)=0$
and $C_{\sigma}(E,\theta)=1-\left|b^N_{\sigma}(E,\theta)\right|^2$
in Eqs. (\ref{3D q}) and (\ref{3D s}), where
$b^N_{\sigma}(E,\theta)$ is given by Eq.~(\ref{bN}). Note that
amplitudes of the conductance oscillations in metallic FSF
junctions ($Z=0$) are few orders of magnitude greater than in the
corresponding FNF junctions. An important consequence of the
coherency is the nontrivial spin conductance for the AP alignment,
which approaches the normal value $G^N_s=0$ either for
$E/\Delta\gg 1$ or in the tunnel limit for all energies.

Incoherent transport through an FSF double junction is described
as a transport through the corresponding FS and SF junctions in
series. In that case, the conductance spectra are calculated using
the generalized BTK probabilities, obtained from Eqs. (\ref{a FS})
and (\ref{b FS}). Numerical results for the incoherent transport
are also presented in Figs.~\ref{GZ0} and \ref{GZ1} for
comparison. It is evident that in thick films the only difference
comes from the interference-effect oscillations for the energies
above the gap. In contrast with the coherent transport,
$G_s(E)\equiv 0$ for the AP alignment, and nonequilibrium spin
density accumulation changes the chemical potential of two spin
subbands in the superconductor. This reduces the superconducting
gap with increasing voltage and destroys the superconductivity at
a critical voltage on the order of $\Delta/e$.\cite{Takahashi}


\section{Conclusion}

We have analyzed coherent electronic transport properties of an
FSF double-barrier junction taking into account the influence of
the exchange interaction, the resistance of the interfaces, and
the Fermi velocity mismatch on the scattering probabilities and
conductance spectra. The exchange potential, the insulating
barriers at the interfaces, and the Fermi velocity mismatch reduce
the Andreev reflection due to the enhancement of normal
reflection. It is shown that subgap electronic transmission and
oscillations of differential conductances are the main features of
the coherent quantum transport through a superconducting layer in
FSF (and NSN) double junctions. In metallic junctions, the subgap
transmission suppresses the excess current through thin
superconducting films. The scattering probabilities and
conductances oscillate as a function of the layer thickness and of
the quasiparticle energy above the gap. Periodic vanishing of the
Andreev reflection (and the excess current) at the energies of
geometrical resonances is found as an important consequence of the
quasiparticle interference. In principle, oscillations of
differential conductances with the period of geometrical
resonances could be used for reliable spectroscopy of
quasiparticle excitations in superconductors.\cite{Nesher}

Finite-size effects, along with the difference between coherent
and incoherent transport, are essential for spin currents in FSF
junctions. Besides the spin-polarized subgap tunneling in thin
superconducting films, pronounced oscillations of spin conductance
are found in thick films. As a consequence of the interference, a
nontrivial spin current without the excess spin accumulation and
without the destruction of superconductivity by voltage is found
even for the antiparallel alignment of the electrode
magnetizations.

\section{Acknowledgment}

We are grateful to Ivan Bo\v{z}ovi\'c for pointing out the
significance of the problem treated in this paper and for help at
the initial stage of this work. Furthermore, we thank Irena
Kne\v{z}evi\'c for useful discussions. This work has been
supported by the Serbian Ministry of Science, project ${\rm
N}^{\circ} 1899$.

\section*{APPENDIX}

The wave functions $\psi_\sigma(z)$, Eqs.
(\ref{psiL})--(\ref{psiR}), satisfy the boundary conditions
\begin{eqnarray}
\label{bc1}
\psi_\sigma (z)|_{z=0_-}&=&\psi_\sigma (z)|_{z=0_+},\\
\frac{d\psi_\sigma (z)}{dz}\Big|_{z=0_-}&=&\frac{d\psi_\sigma
(z)}{dz}\Big|_{z=0_+} -\frac{2m\hat{W}}{\hbar ^2}\psi_\sigma
(0),\\
\psi_\sigma (z)|_{z=l_-}&=&\psi_\sigma (z)|_{z=l_+},\\
\frac{d\psi_\sigma (z)}{dz}\Big|_{z=l_-}&=&\frac{d\psi_\sigma
(z)}{dz}\Big|_{z=l_+}-\frac{2m\hat{W}}{\hbar ^2}\psi_\sigma (l)
\label{bc4}.
\end{eqnarray}
Neglecting $E/E^{(F)}_F$ and $\Delta/E^{(S)}_F$ in the wave
vectors, except in the exponents, Eq.~(\ref{zeta}), solutions of
Eqs.~(\ref{bc1})--(\ref{bc4}) for AR, NR, TE, and TH amplitudes
can be written in the form
\begin{eqnarray}
\label{a general} a_\sigma(E,\theta)&=&\frac{4
(\tilde{k}_{\sigma}/\tilde{q}_\sigma) \Delta
\sin(\zeta_-/2)}{\Gamma}\left[{\cal A}^R_+ E \sin(\zeta_-/2)+i
{\cal B}^R_+ \Omega \cos(\zeta_-/2)\right], \\ \label{b general}
b_\sigma(E,\theta)&=&\frac{1}{\Gamma}[{\cal A}^R_+{\cal
C}_+\Delta^2 - \left({\cal A}^R_+{\cal C}_+E^2 + {\cal B}^R_+{\cal
D}_+\Omega^2 \right)\cos(\zeta_-) + \left({\cal A}^R_-{\cal C}_- +
{\cal B}^R_-{\cal D}_-\right)\Omega^2\cos(\zeta_+) \nonumber\\ &~&
+ i\left({\cal B}^R_+{\cal C}_+ + {\cal A}^R_+{\cal
D}_+\right)E\Omega\sin(\zeta_-) - i\left({\cal B}^R_-{\cal C}_- +
{\cal A}^R_-{\cal D}_-\right)\Omega^2\sin(\zeta_+)], \\ \label{c
general} c_\sigma(E,\theta)&=&\frac{4
(\tilde{k}_{\sigma}/\tilde{q}_\sigma) \Omega
e^{-ik^+_{\bar{\sigma}}l } }{\Gamma}\times \nonumber\\
&~&\times\{ i\left[{\cal F}_+ \cos(\zeta_+/2)+i {\cal E}_+
\sin(\zeta_+/2)\right]E\sin(\zeta_-/2) - \left[{\cal
E}_+\cos(\zeta_+/2)+i {\cal F}_+\sin(\zeta_+/2)
\right]\Omega\cos(\zeta_-/2)\},
\\ \label{d general} d_\sigma(E,\theta)&=&\frac{4
(\tilde{k}_{\sigma}/\tilde{q}_\sigma) \Delta\Omega
e^{ik^-_{\sigma}l}}{\Gamma}\times \nonumber\\ &~&\times
i\left[{\cal F}_- \cos(\zeta_+/2)+i {\cal
E}_-\sin(\zeta_+/2)\right]\sin(\zeta_-/2),
\end{eqnarray}
where
\begin{eqnarray}
\label{Gamma} {\Gamma}={\cal A}^L_+{\cal A}^R_+\Delta^2 -
\left({\cal A}^L_+{\cal A}^R_+E^2 + {\cal B}^L_+{\cal
B}^R_+\Omega^2 \right)\cos(\zeta_-) + \left({\cal A}^L_-{\cal
A}^R_- + {\cal B}^L_-{\cal B}^R_-\right)\Omega^2\cos(\zeta_+)
\nonumber\\ + i\left({\cal A}^L_+{\cal B}^R_+ + {\cal B}^L_+{\cal
A}^R_+\right)E\Omega\sin(\zeta_-) - i\left({\cal A}^L_-{\cal
B}^R_- + {\cal B}^L_-{\cal A}^R_-\right)\Omega^2\sin(\zeta_+).
\end{eqnarray}
In Eqs.~(\ref{a general})--(\ref{Gamma}),
\begin{eqnarray*}
{\cal A}^{L(R)}_{\pm}&=& K^{L(R)}_1 \pm K^{L(R)}_2, \\
{\cal B}^{L(R)}_{\pm}&=& 1 \pm  K^{L(R)}_1 K^{L(R)}_2,
\\ {\cal C}_{\pm}&=& {K^{L}_1}^* \mp K^{L}_2, \\ {\cal
D}_{\pm}&=& -(1 \mp {K^{L}_1}^* K^{L}_2), \\ {\cal E}_{\pm}&=&
K^{L}_2 \pm K^{R}_2, \\ {\cal F}_{\pm}&=& 1 \pm K^{L}_2 K^{R}_2,
\end{eqnarray*}
with
\begin{eqnarray*}
K^L_1&=& \frac{\tilde{k}_{\sigma}+iZ}{\tilde{q}_\sigma},
\\ K^L_2&=&
\frac{\tilde{k}_{\bar{\sigma}}-iZ}{\tilde{q}_\sigma},
\\ \label{Q1R} K^R_1&=&
\frac{\tilde{k}_{\sigma[{\bar{\sigma}}]}+iZ}{\tilde{q}_\sigma}, \\
K^R_2
&=&\frac{\tilde{k}_{{\bar{\sigma}}[\sigma]}-iZ}{\tilde{q}_\sigma},
\end{eqnarray*}
for the P [AP] alignment. Here,
${K^{L}_1}^*=(\tilde{k}_{\sigma}-iZ)/\tilde{q}_\sigma$ is the
complex conjugate of $K^L_1$. General solutions for NSN
double-barrier junctions correspond to $X=0$.

In the corresponding FNF double junction, AR and TH processes are
absent, $A_\sigma=D_\sigma\equiv 0$, and the expression for NR
amplitude, Eq.~(\ref{b general}), reduces to
\begin{equation}
\label{bN} b^N_{\sigma}(E,\theta)=\frac{ ({K^{L}_1}^* -
K^{R}_1)\cos(lq^N_{\sigma})+i (1-{K^{L}_1}^*
K^{R}_1)\sin(lq^N_{\sigma}) }{ (K^{L}_1 +
K^{R}_1)\cos(lq^N_{\sigma})-i(1+K^{L}_1
K^{R}_1)\sin(lq^N_{\sigma}) },
\end{equation}
where $q^N_{\sigma}=\sqrt{(2m/\hbar ^2)(E^{(S)}_F+E)-{\bf
k}^2_{||,\sigma}}$. Setting $\kappa=1$ and $\theta=0$ in Eq.
(\ref{bN}), $1-|b^N_{\sigma}(E,\theta)|^2$ reduces to the result
of Zheng {\it et al.}\cite{Zheng2} for the transmission
coefficient.

To complete our considerations, we also present the probability
amplitudes for an FS single junction in the same notation,
\begin{eqnarray}
\label{a FS} a_\sigma(E,\theta)&=&\frac{2
(\tilde{k}_{\sigma}/\tilde{q}_\sigma) \Delta }{{\cal A}^L_+E +
{\cal B}^L_+\Omega},\\ \label{b FS}
b_\sigma(E,\theta)&=&\frac{{\cal C}_+E + {\cal
D}_+\Omega}{{\cal A}^L_+E + {\cal B}^L_+\Omega},\\
\label{c FS} c_\sigma(E,\theta)&=&\frac{2
(\tilde{k}_{\sigma}/\tilde{q}_\sigma) E\bar{u}(1+K^L_2) }{{\cal
A}^L_+E + {\cal B}^L_+\Omega},\\ \label{d FS}
d_\sigma(E,\theta)&=&\frac{2 (\tilde{k}_{\sigma}/\tilde{q}_\sigma)
E\bar{v}(1-K^L_2)}{{\cal A}^L_+E + {\cal B}^L_+\Omega}.
\end{eqnarray}
Note that $c_\sigma$ and $d_\sigma$ now describe the transmission
of the Bogoliubov electronlike and holelike quasiparticle,
respectively. For $\kappa=1$ and $\theta=0$, Eqs.~(\ref{a FS}) and
(\ref{b FS}) reduce to the results of Zheng {\it et
al}.\cite{Zheng} The well-known BTK results can be reproduced by
taking $X=0$, $\kappa=1$, and $\theta=0$ in Eqs.~(\ref{a
FS})--(\ref{d FS}).\cite{foot}

\clearpage




\begin{thebibliography}{99}
\bibitem{foot} The parameter $Z$ in our definition
differs from the BTK by a factor of 2.
\end{thebibliography}
\end{document}